\documentclass{article}
\usepackage{amsmath,amsfonts,amsthm,amssymb,amscd,color,xcolor,mathrsfs}
\usepackage{amsfonts}
\usepackage{amsmath,amsthm}
\usepackage{hyperref}
\usepackage{latexsym}
\usepackage{array}
\usepackage{amssymb}
\usepackage{enumerate}

\usepackage[francais,english]{babel}

\usepackage{color}
\usepackage[latin1]{inputenc}

\DeclareFontFamily{U}{wncy}{}
    \DeclareFontShape{U}{wncy}{m}{n}{<->wncyr10}{}
    \DeclareSymbolFont{mcy}{U}{wncy}{m}{n}
    \DeclareMathSymbol{\Sh}{\mathord}{mcy}{"58}

\theoremstyle{plain}
\newtheorem{theorem}{Theorem}[section]
\newtheorem*{theorem*}{Theorem}
\newtheorem{lemma}[theorem]{Lemma}

\newtheorem{corollary}[theorem]{Corollary}

\theoremstyle{remark}

\newtheorem{example}[theorem]{Example}

\newtheorem*{lem*}{Lemma}
\newtheorem*{sublem*}{Sublemma}
\newtheorem*{remark*}{Remark}
\newtheorem*{NB*}{NB}

\newcommand{\Sv}{ \Sigma^{nbh}}
\newcommand{\Svm}{ \Sigma_*}
\newcommand{\Sod}{ \Sigma^1}
\newcommand{\nbh}{\text{neighbourhood}}

\newcommand{\nt}{ \lan t \ran }

\newcommand{\R}{ \mathbb{R} }

\newcommand{\Z}{ \mathbb{Z} }

\newcommand{\cC}{ \mathcal{C} }

\newcommand{\cN}{ \mathcal{N} }

\newcommand{\om}{ \omega }

\newcommand{\G}{ \Gamma }

\renewcommand{\phi}{ \varphi }
\newcommand{\eps}{\varepsilon}

\newcommand{\de}{ \delta }

\renewcommand{\a}{ \alpha }

\newcommand{\la}{ \lambda }

\newcommand{\be}{\begin{equation}}
\newcommand{\ee}{\end{equation}}
\newcommand{\ben}{\begin{equation*}}
\newcommand{\een}{\end{equation*}}

\numberwithin{equation}{section}

\newcommand{\lan}{ \langle }

\newcommand{\ran}{ \rangle}

\newcommand{\p}{ \partial}

\author{Sergei Kuksin\footnote{CNRS, Institut de Math\'emathiques de Jussieu--Paris Rive Gauche, UMR 7586, Universit\'e Paris Diderot, Sorbonne Paris Cit\'e, F-75013, Paris, France \& School of Mathematics, Shandong University,  Shanda Nanlu, 27, 250100, PRC; e-mail: \href{mailto:Sergei.Kuksin@imj-prg.fr}{sergei.kuksin@imj-prg.fr}} }

\title
{Asymptotic  properties of integrals of quotients, 
 when  the numerator oscillates and denominator
degenerates}

\date{}
\begin{document}

\maketitle

\hfill  {\it Dedicated to V.A.~Marchenko  on the occasion of his 95-th birthday.}

\begin{abstract}
We study asymptotical expansion as $\nu\to0$ for integrals over ${ \mathbb{R} }^{2d}=\{(x,y)\}$ of quotients of the form 
$F(x,y) \cos(\la x\cdot y) \big/ \big(  (x\cdot y)^2+\nu^2\big)$, where $\la\ge 0$   and $F$ decays 
at infinity sufficiently fast.  Integrals of this kind appear in the theory of wave turbulence. 
 \end{abstract}


\section{Introduction}\label{s2}

In the paper \cite{K} we study asymptotical, as $\nu\to0$, behaviour of integrals 
$$
I_\nu=
 \int_{\R^d\times\R^d}  dx\,dy\, \,   \frac{ F(x,y)}{ (x\cdot y)^2 +(\nu \Gamma(x, y))^2}\,,\quad d\ge2\,,
 0<\nu\ll	1,
$$
where $F$ and $\Gamma$ are   $C^2$-function,  $\Gamma$ is positive and the two satisfy certain conditions at infinity.
In particular, if  $\Gamma\equiv1$, then
\be\label{4}
 |\p_z^\a F(z)|\le C' \lan z\ran^{-N-|\alpha|}\;\;\;\; \forall z=(x,y)\in\R^{2d},\; \forall\, |\a|\le2\,,
\ee
where   $C'>0$ and $N>2d-2$. 
Denote by 
\be\label{sigma}
\Sigma\subset \R^{2d}= \R^d_x\times \R^d_y
\ee
the quadric $\{(x,y): x\cdot y=0\}$, and by $\Sigma_*$ its regular part $\Sigma\setminus\{(0,0)\}$. 
It is proved in  \cite{K}  that 
\be\label{2}
I_\nu =\pi \nu^{-1} \int_{\Svm} \frac{F(z)}{|z|\, \G(z)}\,d_{\Svm} z+O(\chi_d(\nu)), 
\ee
where 
$$
 \chi_d(\nu)=
   \left\{\begin{array}{ll}
 1,& d\ge3 \,,
 \\
 \max\big(\ln(\nu^{-1}),1\big),  & d=2\,,
\end{array}\right.
$$
$d_{\Svm} $ is the volume element on $\Svm$, induced from the standard Riemann structure in $\R^{2d}$, 
and the integral in \eqref{2} converges absolutely.  Integrals of 
this form appear in the study of the 4-waves interaction. The wave turbulence (WT) limit in systems with the 4-waves
interaction leads to oscillating versions of the integrals above with 
 constant functions $\G$. Re-denoting $\nu\G$ back to $\nu$ we write the integrals in question as 
\be\label{3}
\begin{split}
J_\nu=
 \int_{\R^{2d}}  dz\, \,   \frac{ F(z)   \cos(\la x\cdot y)}{ (x\cdot y)^2 +\nu^2}\,,
 \quad d\ge2\,,\;\;
 \la\ge0\,,\;  0<\nu\le1
\end{split}
\ee
(as before, $z=(x,y)$). 
We assume that $F$ is a $C^2$--function, satisfying \eqref4.

The aim of this work is to prove the following  result, describing the asymptotical behaviour of 
$J_\nu$
when $\nu\to0$, uniformly in $\lambda\ge0$:

\begin{theorem}\label{t_1}
Let  $0<\nu\le1$  and  $ \la\ge0$. Then the integral $J_\nu$ and the integral 
$$
J_0 = \pi e^{-\nu\lambda}
 \int_{\Svm} {F(z)}{|z|^{-1} }\,d_{\Svm} z 
$$
converge absolutely and 
\be\label{5}
\big| J_\nu  -  \nu^{-1} J_0\big| \le C  \chi_d(\nu), 
\ee
where  $C$ depends on $d$ and the constants $C', N$  in \eqref{4},
but not on  $\nu$ and  $\la$.
\end{theorem}

Note that since  $C$ does not depend on $\lambda$, then relation \eqref{5} remains valid for integrals 
\eqref{3}, where $\lambda=\lambda(\nu)$ is any function of $\nu$. Concerning the imposed restriction
$d\ge2$ see item~iv) in Section~\ref{s_5}. 

If $\la=0$, the integral $J_\nu$ becomes a special case of  $I_\nu$ (with $\G=1$), and \eqref{5} follows
from \eqref{2}. Since $\sin^2(\frac{\la}2 x\cdot y) = \tfrac12 (1-\cos(\la x\cdot y))$, then combining \eqref{2} and \eqref{5} we get 

\begin{corollary}
As $\nu\to0$, 
\be\label{55}
\begin{split}
 \int_{\R^d\times\R^d} & dx\,dy\, \,   \frac{ F(x,y)   \sin^2(\frac{\la}2  x\cdot y)}{ (x\cdot y)^2 +\nu^2}\\
& =\frac{\pi}2\, \nu^{-1}   (1- e^{-\nu\la})
 \int_{\Svm} \frac{F(z)}{|z|}\,d_{\Svm} z+O(\chi_d(\nu))\,,
\end{split}
\ee
uniformly in $\lambda\ge0$. 
\end{corollary}

Classically the WT considers  singular versions of the integral in the l.h.s. of \eqref{55}:
\be\label{WT}
\int dx\,dy\,  \frac { F(x,y)   \sin^2(\tfrac{\la}2  x\cdot y)}{ (x\cdot y)^{2} }\,.
\ee
The theory deals with these integrals by 
performing  certain  formal calculations, 
see Section~6 of \cite{Naz} (e.g. note there eq.~(6.39)+(6.41)). Assertion \eqref{55} may be
regarded as a regularisation of the integral \eqref{WT}. The factor $|z|^{-1}$ which it introduces 
in the limiting density  is not present in the asymptotical description of integrals \eqref{WT},
used in the works on WT.

Theorem \ref{t_1} is proved below in Sections~\ref{s_2}--\ref{s_4}, using 
the geometric approach of the paper \cite{K}, which also applies to various 
modifications of integrals $I_\nu$ and $J_\nu$. Some of these applications 
are discussed in the last Section~\ref{s_5}. 

 \medskip
 
\noindent 
{\bf Notation.}
As usual, we denote $\lan z\ran = \sqrt{|z|^2+1}$.
For an integral $I=\int_{\R^{2d}} f(z)\,dz$  and a submanifold $M\subset\R^{2d}$, dim$\,M=m\le 2d$,
compact or not (if $m=2d$, then $M$ is an open domain in $\R^{2d}$) we write
$
\lan I,M\ran =  \int_Mf(z)\,d_M(z), 
$
where $d_M(z)$ is the volume--element on $M$, induced from $\R^{2d}$. Similar $\lan |I|,M\ran$
stands for the integral $\int_M |f(z)|\,d_M(z)$.
\smallskip

\noindent 
{\bf Acknowledgments.}  We acknowledge the support from the Centre National de la Recherche Scientifique (France)
 through the grant PRC CNRS/RFBR 2017-2019 No~1556 
 ``Multi-dimensional semi-classical problems of condensed matter physics and quantum dynamics''.

\section
{Geometry of the quadric 
$\{x\cdot y=0\}$ and its vicinity.}
\label{s_2}

\subsection
{The geometry of the quadric.}

The difficulty in studying  the integral $J_\nu$ with small $\nu$ comes from the vicinity of the quadric 
$\Sigma=\{x\cdot y=0\}$. 
To examine  the integral's behaviour  there we first analyse the geometry of the vicinity of the 
regular part of the quadric  $\Sigma_*=\Sigma\setminus\{(0,0)\}$, following \cite{K}. Example~\ref{ex} at the end of the 
paper provides an elementary illustration to the objects, involved in this analysis. 

The set  $\Sigma_*$   is a smooth submanifold of $\R^{2d}$ of dimension $2d-1$. 
We denote by $\xi$ 
a local coordinate on $\Sigma_*$ with the coordinate mapping
$\xi\mapsto (x_\xi,y_\xi)=z_\xi\in \Sigma_*$, denote 
 $|\xi|= |(x_\xi,y_\xi)|$ and denote 
$
N(\xi)= (y_\xi, x_\xi).
$ The latter is 
the  normal to $\Sigma_*$ at $\xi$ of  length $|\xi|$.
For any $0\le R_1<R_2$ we set 
\begin{equation*}
\begin{split}
&S^{R_1} = \{z\in\R^{2d}: |z|=R_1\}\,,\quad  \Sigma^{R_1} = \Sigma\cap S^{R_1} \,,
\\
& S^{R_2}_{R_1} = \{z: R_1< |z|<R_2\}\,,\quad \Sigma^{R_2}_{R_1}  = \Sigma\cap S^{R_2}_{R_1}\,,
\end{split}
\end{equation*}
and for  $t>0$  denote by $D_t$ the dilation operator
$$
D_t:  \R^{2d}\to \R^{2d},\quad z \mapsto tz\,.
$$ 
For $z=(x,y)$ we write $\omega(z) = x\cdot y$. 

The following result is Lemma 3.1 from \cite{K}:

\begin{lemma}\label{l_p1}
1)
There exists $\theta_0\in(0, 1]$
such that  a suitable \nbh\  $\Sv=\Sv(\theta_0)$ of $\Sigma_*$ in $\R^{2d}\setminus\{0\}$, is invariant with respect 
to the dilations $D_t$, $t>0$, and may be uniquely parametrized  as
$$
\Sv=
\{ \pi(\xi,\theta): 
\xi\in\Sigma_*,\; |\theta|<\theta_0\}\,,
$$
where $\pi(\xi,\theta) = (x_\xi,y_\xi) +\theta N_\xi = (x_\xi,y_\xi) + \theta(y_\xi, x_\xi) $. In particular,
$
|\pi(\xi,\theta)|^2= |\xi|^2(1+\theta^2)
$.\\
2) If $\pi(\xi,\theta)\in\Sv$, then 
 \be\label{p0}
 \omega \big( \pi(\xi,\theta)\big) =|\xi|^2\theta.
  \ee
  3) If $(x,y)\in S^R\setminus \Sv$, then $|x\cdot y| \ge cR^2$ for some $c=c(\theta_0)>0$. 
\end{lemma}

For  $0\le R_1<R_2$ we will denote 
$$
(\Sv)_{R_1}^{R_2} = \pi (\Sigma_{R_1}^{R_2} \times \big(-\theta_0, \theta_0)\big). 
$$
Now we discuss the Riemann geometry of the domain $\Sv=\Sv(\theta_0)$,
following \cite{K}.

The set  $\Sigma$ is a cone with the vertex in the origin,  and 
$
\Svm = \{ tz: t>0, z\in \Sigma^1\}
$.
The set  $\Sigma^1$ is a closed manifold of dimension $2d-2$.  Let us cover it by a finite system of charts
   $\cN_1,\dots, \cN_{\tilde n}$, $\cN_j =\{\eta^j = (\eta_1^j,\dots,\eta^j_{2d-2})\}$, and for any chart $\cN_j$
    denote by $m(d\eta^j)$ the volume element    on $\Sigma^1$, induced from $\R^{2d}$. Below
    we write points in any chart as $\eta$, and the volume element -- as $m(d\eta)$.
   
      The mapping 
   $$
   \Sod\times \R^+ \to \Sigma_*,\quad ((x_\eta,y_\eta), t)\to D_t (x_{\eta},y_{\eta})
   $$
   is a  diffeomorphism. Accordingly,     we can cover $\Sigma_*$ by the $\tilde n$
   charts $\cN_j\times \R_+$ with the coordinates
   $
   (\eta^j, t)=: (\eta,t).
      $
  The coordinates $(\eta,t,\theta)$, where $\eta \in\cN_j, t>0$ and $ |\theta|<\theta_0$,  $1\le j\le\tilde n$,  
     make  coordinate systems on
   $\Sv=\Sv(\theta_0)$. In the coordinates $(\eta,t)$ 
    the volume element on $\Sigma_*$ is 
   \be\label{vol_on_*}
   d_{\Svm} =
   t^{2d-2}m(d\eta)\,dt\,.
   \ee
In the coordinates $(\eta,t,\theta)$ the volume elements in $\R^{2d}$ reeds 
 \be\label{p4}
 dx\,dy =t^{2d-1} \mu(\eta,\theta) m(d\eta)dt\,d\theta\,, \quad \text{where}\;
 \mu(\eta,0)=1\,
 \ee
 (see \cite{K}),    a dilation map $D_r$,  $r>0$, reeds
 $
D_r (\eta, t,\theta)  = (\eta, rt, \theta)\,,
$
and by \eqref{p0} 
\be\label{p010}
\omega(\eta, t, \theta) = t^2\theta\,.
\ee
Finally, since 
$
\frac{\p}{\p \theta} = \nabla_z\cdot(y,x), 
$
then  in view of \eqref{4}  for any $(\eta, t, \theta)$ and any $k\le2$ 
  \be\label{new_est}
  \big| \frac{\p^k}{\p \theta^k} F(\eta, t, \theta) \big|  \le C  \nt^{-N}\,, \quad\, N>2d-4.
  \ee

\subsection
{The volume element $d_{\Svm} $ and the measure $|z|^{-1}d_{\Svm}. $
 }
 
 Theorem \ref{t_1} and the results of \cite{K} shows that the manifold $\Sigma_*$, equipped with the volume $d_{\Svm}$
 and the   measure $|z|^{-1}d_{\Svm}$, 
 is crucial to study asymptotic of integrals $I_\nu$, $J_\nu$ and their similarities (cf. Section~6 of \cite{K} and Section~5 below). The coordinates  $(\eta, t)$ and the presentation \eqref{vol_on_*} for the volume element are sufficient for the purposes of this 
 work. But the quadric $\Sigma$ is reach in structures and admits more instrumental coordinate systems. In particular, 
 if $d=2$ we can introduce in the space $\R^2_x$ in \eqref{sigma} the polar coordinates $(r,\phi)$. Then for any fixed 
 non-zero vector  $x=(r,\phi)\in \R^2_x$  the set $\{y\in\R^2_y: (x,y)\in \Sigma_*\}$ is the line in $\R^2_y$, perpendicular to $x$,
 and 
 having the angle $\phi+\pi/2$ with the horizontal axis. Parametrizing it by the length-coordinate $l$ we get on $\Sigma_*$
 the coordinates $(r,l,\phi) \in \R^+\times \R \times S^1$, $S^1= \R/ 2\pi\Z$, with the coordinate mapping 
 $$
 \Phi: (r,l,\phi) \mapsto \big( x=(r\cos\phi, r\sin\phi), y=(-l\sin\phi, l\cos\phi)\big)
 $$
(this map is singular at $r=0$). Since 
\begin{equation*}
\begin{split}
&|\p\Phi/ \p r|^2 =1, \;  |\p\Phi/ \p l |^2 =1, \; |\p\Phi/ \p \phi|^2 = r^2+l^2,\\
& \lan \p\Phi/ \p r, \p\Phi/ \p l \ran=  \lan \p\Phi/ \p r, \p\Phi/ \p \phi \ran=  \lan \p\Phi/ \p l, \p\Phi/ \p \phi \ran=0,
\end{split}
\end{equation*}
then in these coordinates the volume element on $\Sigma_*$ reeds as $\sqrt{r^2+l^2} dr\,dl\,d\phi$,
and the  measure $|z|^{-1}d_{\Svm}$ -- as $ dr\,dl\,d\phi$. Consider the fibering 
$$
\Pi: \R^2_x \times \R^2_y \supset \Sigma_* \to \R^2_x,\quad (x,y)\mapsto x. 
$$
It has a singular fiber $\Pi^{-1}0 = \{0\}\times \R^2_y$, and for any non-zero $x$ the fiber $\Pi^{-1}x$ equals
$ \{x\}\times x^\perp $, where $x^\perp$ is the line in $\R^2_y$, perpendicular to $x$. Since 
$dx= rdr\,d\phi$, then the given above presentation for the measure $|z|^{-1}d_{\Svm}$ implies that its 
restriction to the regular part $\Sigma_*^+$ 
 of the fibered manifold $\Sigma_*$,  $\ \Sigma_*^+= \Sigma_*\setminus (\{0\}\times \R^2_y)$, 
disintegrates by the foliation $\Pi$ as 
\be\label{f1}
(|z|^{-1}d_{\Svm})\mid_{\Sigma_*^+} = |x|^{-1} dx\,d_{x^\perp} y,\quad x\ne0,\ y\in x^\perp,
\ee
where $d_{x^\perp}$ is the length on the euclidean line $x^\perp\subset \R^2_y$. 

We do not undertake  the job of getting a right   analogy of this result  for  the multidimensional 
case $d>2$, but note that a straightforward  modification of the construction above leads to the observation that 
for any $d\ge2$ the measure   $|z|^{-1}d_{\Svm}$, restricted to $\Sigma_*^+$, disintegrates as 
\be\label{f2}
p_d(x,y) dx\, d_{x^\perp} y,\quad x\in\R^d\setminus\{0\},\ y\in x^\perp,
\ee
where $x^\perp$ is the orthogonal complement to $x$ in $R^d_y$, 
$,d_{x^\perp}$ is the volume element on this euclidean space, and the function $p_d$ satisfies the estimate
$ p_d \le C (|x|+|y|)^{d-2} |x|^{1-d}$.

 \section{ Integral over the vicinity of $\Sigma$  }
 
 To study the behaviour 
of the integral over a  \nbh\ of $\Sigma$  we first prove that the integral, evaluated over the vicinity 
  of the singular point  $(0,0)$ is small, and next  study  the integral over the vicinity
of the regular part $\Sigma_*$ of the quadric.

For $ 0<\delta\le1$ denote
$$
K_\delta = \{ z=(x,y):|x|\le\delta, |y|\le\delta\}\subset \R^d\times \R^d \,.
$$
An upper bound for the integral over $K_\delta $ follows from Lemma~2.1 of \cite{K}:
\be\label{2.1}
|\lan |J_\nu|, K_\de\ran | \le 
    \int_{K_\delta} \frac{ |F(z)|\,dz}{(x\cdot y)^2 +\nu^2} \le 
 {C} {\nu^{-1} \delta^{2d-2}} \,. 
\ee

Now we estimate the integral over the \nbh\  $\Sv$ of $\Svm$.  For this end, using \eqref{p4}, 
 for $0\le A< B\le\infty$ we disintegrate  
$\lan J_\nu, (\Sv)_A^B\ran$ as 
 \be\label{3.9}
 \begin{split}
 \lan J_\nu,  (\Sv)_A^B\ran=
 \int_{\Sod} m(d\eta) \int_{A}^{B} dt\,  t^{2d-1}   \int_{-\theta_0}^{\theta_0} d\theta\,
 \frac{F(\eta, t,\theta) \mu(\eta,\theta) \cos(\la x\cdot y) }{ t^4 \theta^2 +\nu^2}\\
=  \int_{\Sod} m(d\eta) \int_{A}^{B} dt \, t^{2d-1} \Upsilon_\nu(\eta,t)\,,
 \end{split}
 \ee
 where 
 $$
 \Upsilon_\nu(\eta,t) 
 = t^{-4} \int_{-\theta_0}^{\theta_0} 
  \frac{F(\eta,t,\theta)\mu(\eta,\theta)   \cos(\la t^2\theta)  \,d\theta}{\theta^2  +\eps^2}\,, \quad \eps =   \nu t^{-2}\,.
 $$

To study   $\Upsilon_\nu$ we first 
consider  the integral $\Upsilon_\nu^{0}$,
 obtained from $\Upsilon_\nu$ by frozening $F\!\mu$ at $\theta=0$. Since $\mu(\eta, 0)=1$, then 
$$
 \Upsilon_\nu^{0}= 2 t^{-4}    F (\eta,t,0) \int _{0}^{\theta_0}\frac{  \cos(\la t^2\theta) \,d\,\theta}{\theta^2+\eps^2}
 =2 \nu^{-1} t^{-2}    {F (\eta,t,0)}  \int _{0}^{\theta_0/\eps}\frac{  \cos( \nu\la w) \,dw}{w^2+1}\,.
  $$
  Consider the integral 
  $$
 2  \int _{0}^{\theta_0/\eps}\frac{  \cos(\nu\la  w) \,dw}{w^2+1} =  2  \int _{0}^{\infty }\frac{  \cos(\nu\la  w) \,dw}{w^2+1}
 - 2  \int _{\theta_0/\eps}^{\infty}\frac{  \cos(\nu\la  w) \,dw}{w^2+1} =: I_1 - I_2\,. 
  $$
  Since
  $$
   2  \int _{0}^{\infty}\frac{  \cos(\xi  w) \,dw}{w^2+1} =   \int _{-\infty}^{\infty}\frac{  e^{i\xi  w} \,dw}{w^2+1} = \pi e^{-|\xi|},
  $$
  then $I_1 = \pi e^{-\nu\la}$. For $I_2$ we have an obvious bound 
  $
  |I_2| \le 2\eps/\theta_0  = C_1\nu t^{-2}  \,.
  $
   So
  \be\label{4.9}
  \begin{split}
  \Upsilon_\nu^0 (\eta,t)=\pi\nu^{-1}   t^{-2}  F (\eta,t,0) (e^{-\nu\la} +\Delta_t)\,, \qquad
  |\Delta_t | \le C \nu t^{-2}  \,.
  \end{split}
  \ee

  Now we estimate the difference between $\Upsilon_\nu$ and $\Upsilon_\nu^{0}$. Writing 
  $(F\mu)(\eta, t,\theta) - (F\mu)(\eta, t,0)$ as $A(\eta, t)\theta + B(\eta, t,\theta)\theta^2$, where 
  $ |A|, |B| \le C \nt^{-N}$ in view of \eqref{new_est}, we have 
  $$
  \Upsilon_\nu - \Upsilon_\nu^0 =   t^{-4} \int_{-\theta_0}^{\theta_0} 
  \frac{(A\theta +B \theta^2 )  \cos(\la t^2\theta)  \,d\theta}{\theta^2  +\eps^2}\,.
  $$
  Since the first integrand is odd in $\theta$, then its integral vanishes, and 
  $$
 | \Upsilon_\nu - \Upsilon_\nu^0| \le C \nt^{-N}   t^{-4} \int_{-\theta_0}^{\theta_0} 
  \frac{ \theta^2    \,d\theta}{\theta^2  +\eps^2}\le 2 C \nt^{-N}   t^{-4}\theta_0 \,.
  $$
  So by \eqref{4.9}
   \be\label{4.11}
  \begin{split}
 |\Upsilon_\nu (\eta,t) &-  \pi \nu^{-1}   t^{-2}  F (\eta,t,0) e^{-\nu\la}|\\
 &\le  C\nt^{-N}\big( t^{-4} +\nu^{-1} t^{-2} \, \nu t^{-2}\big)  \le C' \nt^{-N} t^{-4} \,.
  \end{split}
  \ee

    \section{End of the proof of Theorem \ref{t_1}} \label{s_4}
    
    \noindent 
   1) In view of \eqref{3.9}, \eqref{4.11} and  since $N>2d-2$, 
   for $\delta\in(0,1]$ we have 
    \begin{equation*}
    \begin{split}
   \big|  \lan J_\nu, \big(\Sv \big)_\delta^\infty \ran - \pi\nu^{-1}e^{-\nu\la} \int_{\Sigma^1}m\,d\eta\int_\de^\infty dt\,
   t^{2d-3}  F (\eta,t,0)\, \big|\\
    \le C \int_\de^\infty t^{2d-5} \nt^{-N}\,dt \le C_1 \chi_d(\delta)\,.
     \end{split}
      \end{equation*}
      
          \noindent 
   2) Since $d\ge2$ and    $N> 2d-2$, then     by  estimate \eqref{new_est} the integral \\
      $\ 
      \int_{\Sigma^1}m\,d\eta\int_0^\infty dt\, t^{2d-3}    F (\eta,t,0)
      $ 
      converges absolutely, and by \eqref{vol_on_*}  it equals 
      $$
       \int_{\Sigma^1}m\,d\eta\int_0^\infty dt\, t^{2d-3} F(\eta, t, 0)  =  \int_{\Svm} |z|^{-1} F(z) \, d_{\Svm}z\,.
      $$
      
      \noindent 
   3)  Applying 1) and 2) to $F$ replaced by $F_0= C'\langle z\rangle^{-N}$ and using that $|F|\le |F_0|$ by 
   \eqref{4} we find that the integral 
   $\lan J_\nu, \big(\Sv \big)_\delta^\infty \ran $ also converges absolutely. 
   \medskip

     \noindent 
   4) As $|\pi(\xi, \theta)| \le \sqrt2\,|\xi|$, then 
   $
   (\Sv)_0^\delta \subset  S_0^{\sqrt2\delta} \subset K_{\sqrt2\delta}.
   $
   Therefore  by \eqref{2.1} 
        \begin{equation*}
    \begin{split}
   \big|  \lan J_\nu, \big(\Sv \big)_0^\delta \ran - \pi\nu^{-1}e^{-\nu\la} \int_{\Sigma^1}m\,d\eta\int_0^\delta dt\,
   t^{2d-3}   F (\eta,t,0)\, \big|\\
  \le \lan |J_\nu|, K_{\sqrt2\delta} \ran + \pi\nu^{-1}e^{-\nu\la} \int_{\Sigma^1}m\,d\eta\int_0^\delta dt\,
   t^{2d-3}   |F (\eta,t,0)|\\
    \le C_1 \nu^{-1} \de^{2d-2} +C_2 \nu^{-1}\de^2\,.
         \end{split}
      \end{equation*}
      Choosing $\de = \sqrt\nu$, from here and 1)-3) 
      we find that 
        \begin{equation*}
      \begin{split}
   \big|  \lan J_\nu, \Sv  \ran - \pi \nu^{-1}e^{-\nu\la} \int_{\Sigma^1}m\,d\eta\int_0^\infty dt\,
   t^{2d-3}    F (\eta,t,0)\, \big| 
    \le C \chi_d(\nu),
         \end{split}
      \end{equation*}
    and that the integral $ \lan J_\nu, \Sv  \ran$ converges absolutely. 
    \medskip
      
        \noindent 
   5)
      Now let us estimate the integral over $\R^{2d}\setminus \Sv$:
      $$
         \lan |J_\nu|, \R^{2d}\setminus  \Sv  \ran  
       \le  
       \int_{\{ |z| \le \sqrt\nu\}}\frac{|F|\,dz}{\om^2+\nu^2}
       + C_d \int_{\sqrt\nu}^\infty dr\, r^{2d-1}  \int_{S^r\setminus\Sv} \frac{|F(z)|\,d_{S^r}}{\omega^2 + \nu^2}.
      $$
      By item 3) of Lemma \ref{l_p1}, $|\omega|\ge Cr^2$ in ${S^r\setminus\Sv}$. Jointly with \eqref{2.1} this
      implies that 
      $$
          \lan |J_\nu|, \R^{2d}\setminus  \Sv  \ran  \big| \le C + C \int_{\sqrt\nu}^\infty r^{2d-1} r^{-4}\lan r\ran^{-N}
        \, dr \le C_1\chi_d(\nu).
      $$
      So the integral $J_\nu$ converges absolutely and, in view of 4) and 2), 
      \be\label{5.1}
      \begin{split}
       \big|   J_\nu  &-\pi \nu^{-1}e^{-\nu\la} \int_{\Sigma^1}m\,d\eta\int_0^\infty dt\,
   t^{2d-3}    F (\eta,t,0)\, \big| 
\\  
   & = \big|   J_\nu -\pi \nu^{-1}e^{-\nu\la} 
    \int_{\Svm} |z|^{-1} F(z) \, d_{\Svm}z
    \, \big| 
    \le C \chi_d(\nu)\,.
    \end{split}
      \ee
   
    This proves Theorem~\ref{t_1}. 
    
\section{Comments}\label{s_5}

i) The only part of the proof, where we use that $M>2d-2$ is Step~2) in Section~\ref{s_4}:  there this relation is evoked to establish 
the absolute convergence of the integral $J_0$; everywhere else it suffices to assume that $N>2d-4$. Accordingly, if
$F$ satisfies \eqref{4} with $N>2d-4$ and $\langle |F|, \Sigma_1^\infty\rangle<\infty$, then \eqref{5} holds, since 
$\langle |F|, \Sigma_0^1\rangle<\infty$, see Step~4) Section~\ref{s_4}.
\medskip

ii) Our approach does not apply to study integrals \eqref3, where the divisor $(x\cdot y)^2+\nu^2$ is replaced by 
$(x\cdot y)^2+(\nu\Gamma(x,y))^2$ and $\Gamma\ne\,$Const. But it applies to integrals 
$$J^s_\nu=
 \int_{\R^{2d}}  dz\, \,   \frac{ F(z)   \sin (\la x\cdot y)}{ (x\cdot y)^2 +\nu^2}\,,
$$
under certain restrictions on $\lambda$. E.g., if $1\le\lambda\le\nu^{-1}$ and $d\ge3$, then $J^s_\nu=O(1)$ as 
$\nu\to0$, and the leading term again is given by an integral over $\Sigma_*$. The case $d=2$ is a bit  more complicated. 
\medskip

iii) The approach  allows to study integrals \eqref3, where the quadratic form $z\mapsto x\cdot y$ is replaced by any 
non-degenerate indefinite quadratic form of $z\in \R^M$, $M\ge4$.

 iv) The restriction $\,M\ge 4$ in iii)
 (and $d\ge2$ in the main text, where dim$\,z=2d$)  was imposed 
 since near the origin the disparity 
 \eqref{5.1} is controlled by the integral $\int_0 t^{ M-5}\,dt$, which strongly 
 diverges if  $\,M<4$. The difficulty disappears if   $F$ vanishes near zero. This 
  may be illustrated by the following easy example:
 
 \begin{example}\label{ex}
 Consider 
 $$
 J'_\nu =  \int_{\R^2} \frac{F(x,y) \cos(\la xy)}{x^2y^2 +\nu^2}\, dxdy\,,
 $$
 where $F\in C^2_0(\R^2)$ vanishes near the origin. Now $2d=2$, 
 the quadric $\Sigma' = \{xy=0\}$ is one dimensional, 
 has a singularity at the origin and its smooth part ${\Sigma'}^{*} = \Sigma'\setminus {0}$ has four connected components. 
 Consider one of them:
 $
 \cC_1=\{(x,y): y=0, x>0\}.
 $
 Now the coordinate 
 $\xi$  is a point in $\R_+$ with $(x_\xi, y_\xi)= (\xi,0)$ and  with the normal $N(\xi)=(0,\xi)$, 
 the set $\Sigma_1\cap \cC_1$ is the single  point $(1,0)$
 and the coordinate $(\eta,t,\theta)$ in the vicinity of $\cC_1$ degenerates to $(t,\theta)$, $t>0$, $|\theta|<\theta_0$, with the coordinate-map
 $(t,\theta) \mapsto (t, t\theta)$. The relations \eqref{vol_on_*} and \eqref{p4} are now 
 obvious, and the integral \eqref{2.1} vanishes if $\delta>0$ is 
sufficiently small. Interpreting  $z=(x,y)$ as a complex number, we write the assertion of Theorem \ref{t_1} as
$$
\big| J'_\nu  - \pi \nu^{-1} e^{-\nu\la}
 \int_{\Sigma'} \frac{F(z)}{|z| }\,d  z\big| \le C\,,
$$
where  the integral is a contour integral in the complex plane.
 \end{example}

\end{document}